\documentclass[final,5p,times]{elsarticle}

\usepackage{amsmath}
\usepackage{amssymb}
\usepackage{graphicx}

\newcommand{\be}[0]{\begin{equation}}
\newcommand{\ee}[0]{\end{equation}}

\newcommand{\kt}[0]{\text{k}_{\text{T}}}

\def\bnslash{\bar n\!\!\!\slash}

\newcommand{\bn}{{\bar n}}
\newcommand{\mcdot}{\!\cdot\!}
\newcommand{\vect}[1]{\mathbf{#1}}
\newcommand{\abs}[1]{\left\lvert #1\right\rvert}
\newcommand{\bra}[1]{\left\langle #1\right\rvert}
\newcommand{\ket}[1]{\left\lvert #1\right\rangle}
\newcommand{\Lqcd}{\Lambda_{\text{QCD}}}

\newcommand{\eq}[1]{Eq.~\eqref{#1}}
\newcommand{\eqs}[2]{Eqs.~\eqref{#1} and \eqref{#2}}

\renewcommand{\sec}[1]{Sec.~\ref{#1}}

\newcommand{\fig}[1]{Fig.~\ref{#1}}
\newcommand{\tab}[1]{Table~\ref{#1}}
\DeclareMathOperator{\Tr}{Tr}

\newcommand{\CF}{C_F}

\newcommand{\CA}{N_C}

\newcommand{\cO}{\mathcal{O}}

\newcommand{\incl}{ {\rm incl}}

\newcommand{\tabruleA}{\rule{-1pt}{2ex} \rule[-1ex]{0pt}{0pt}}
\newcommand{\tabruleB}{\rule{-2pt}{3.25ex} \rule[-1.5ex]{-2pt}{0pt}}

\begin{document}



\begin{frontmatter}

\title{\textbf{Consistent Factorization of Jet Observables in Exclusive Multijet Cross Sections}}

\author[uw]{Stephen D. Ellis}
\ead{sdellis@uw.edu}

\author[uc]{Andrew Hornig}
\ead{ahornig@berkeley.edu}

\author[uc]{Christopher Lee}
\ead{clee@berkeley.edu}

\author[uw]{Christopher K. Vermilion}
\ead{verm@uw.edu}

\author[uw]{Jonathan R. Walsh}
\ead{jrwalsh@uw.edu}

\address[uw]{University of Washington, Seattle, WA  98195, USA}
\address[uc]{Center for Theoretical Physics, 
University of California,  and Theoretical Physics Group, Lawrence Berkeley National Laboratory,  Berkeley, CA 94720, USA}


\date{\today}


\begin{abstract}

We demonstrate the consistency at the next-to-leading-logarithmic (NLL) level of a factorization theorem based on Soft-Collinear Effective Theory (SCET) for jet shapes  in $e^+e^-$ collisions. We consider measuring jet observables in exclusive multijet final states defined with cone and $\kt$-type jet algorithms. Consistency of the factorization theorem requires that the renormalization group evolution of hard, jet, and soft functions  is such that the physical cross section is independent of the factorization scale $\mu$. The anomalous dimensions of the various factorized pieces, however, depend on the color representation of jets, choice of jet observable, the number of jets whose shapes are measured, and the jet algorithm, making it highly nontrivial to satisfy the consistency condition. We demonstrate the intricate cancellations between anomalous dimensions that occur at the NLL level, so that, up to power corrections that we identify, our factorization of the jet shape distributions is consistent for any number of quark and gluon jets,  for any number of jets whose shapes are measured or unmeasured, for any angular size $R$ of the jets, and for any of the algorithms we consider. Corrections to these results are suppressed by the SCET expansion parameter $\lambda$ (the ratio of soft to collinear or collinear to hard scales) and in the jet separation measure $1/t^2 = \tan^2(R/2)/\tan^2(\psi/2)$, where $\psi$ is the angular separation between jets. Our results can be used to calculate a wide variety of jet observables in multijet final states to NLL accuracy. 

\end{abstract}


\end{frontmatter}

\section{Introduction}

Final states that contain several jets are important Standard Model backgrounds to many new physics processes in high-energy colliders, in addition to serving as sensitive probes of Quantum Chromodynamics (QCD) itself over a wide range of energy scales. The structure of jet-like final states contains signatures of the hard scattering of parton-like degrees of freedom, the branching and showering at ever lower energies, and hadronization at the lowest scale $\Lqcd$. Probing the structure of jets both teaches us about QCD and can help us to distinguish jets of Standard Model origin from those that are truly signatures for new physics. 

The presence of multiple scales governing jets is at once an opportunity to probe many aspects of their physics and also a challenge due to the generation of large logarithms of ratios of these scales spoiling the behavior of perturbation theory. A powerful framework to separate physics at different scales and  to improve the behavior of perturbation series is effective field theory (EFT). EFTs aid in factorizing an observable dependent on multiple scales into pieces  each sensitive to a single energy scale. Renormalization group (RG) evolution of these pieces in  EFT achieves resummation of large logarithms to all orders in perturbation theory. Factorization also allows the disentangling of perturbative and non-perturbative physics \cite{Collins:1989gx,Sterman:1995fz}.

Soft-Collinear Effective Theory (SCET) \cite{Bauer:2000ew,Bauer:2000yr,Bauer:2001ct,Bauer:2001yt} has  had considerable success in applications to many hard-scattering cross sections \cite{Bauer:2002nz} and jet cross sections. SCET separates degrees of freedom in QCD  into distinct soft and collinear modes, expanding the full theory in a parameter $\lambda$ that characterizes the size of collinear momenta transverse to the jet direction, and provides a framework to factorize cross sections into separate pieces coming from interactions at hard, collinear, and soft scales. This was done in SCET for event shape variables using hemisphere jet algorithms in $e^+e^-$ colliders ~\cite{Fleming:2007xt,Hornig:2009vb} and for ``isolated Drell-Yan'' (where central jets are vetoed) in hadron colliders \cite{Stewart:2009yx}. 
In addition, there has been progress in understanding how to implement jet algorithms other than the simple hemisphere jet algorithm in SCET. 
In \cite{Bauer:2003di, Trott:2006bk}, total two-jet rates where the jets are defined by Sterman-Weinberg jet algorithms were computed at NLO. These results were extended to the cases of the exclusive $\kt$ and JADE algorithms in \cite{Cheung:2009sg}. 

In most applications of SCET to exclusive jet cross sections considered to date, there are two back-to-back jets. (Recently Ref.~\cite{Becher:2009th} considered direct photon production in hadron collisions, involving three collinear directions.) In this work we consider for the first time exclusive $N$-jet final states with arbitrary $N\ge 2$ for the SISCone \cite{Salam:2007xv}, Snowmass \cite{Huth:1990mi}, inclusive $\kt$ \cite{Ellis:1993tq}, anti-$\kt$ \cite{Cacciari:2008gp}, and Cambridge-Aachen \cite{Dokshitzer:1997in} jet algorithms. We find that a new feature that arises when more than two jets are present is that the parameter $\lambda$ is not in itself sufficient to ensure factorization. In particular, 
factorization is valid to leading order in $\lambda$ \emph{and} in a jet separation measure $1/t$, where $t$ is defined by
 \be
 t = \frac{\tan(\psi/2)}{\tan(R/2)}
 \,,\ee 
 with $R$ the angular size of a jet as defined by a jet algorithm and $\psi$ the minimum angle between two jets. This is due to the fact that jets need to be both well-collimated ($\lambda \ll 1$) and well-separated ($t \gg 1$). The latter requirement is trivial for back-to-back jets since $1/t = 0$ for $\psi = \pi$.

Our analysis applies not only to the total $N$-jet cross section, but also in the case that 
jet observables are measured on some number $M \le N$ of the jets. We will illustrate the measurement of  angularities $\tau_a$ (cf. \cite{Berger:2003iw,Almeida:2008yp}), defined by
\begin{equation}
\tau_a(J) = \frac{1}{2E_J}\sum_{i\in J} \abs{\vect{p}_T^i} e^{-\eta_i(1-a)}\,,
\end{equation}
where $E_J$ is the energy of the jet $J$, the sum is over particles $i$ in the jet, and $p_T^i$ and $\eta_i$ are the transverse momentum and (pseudo-)rapidity of particle $i$ with respect to the jet axis.  However, most of our results do not depend on this choice of observable, and we organize the calculation such that  other observables can be easily implemented. 
Distributions of jet shapes such as angularities contain logarithms of $\tau_a$ that become large in the limit $\tau_a \to 0$. The factorization theorem we present provides the basis for resummation of these logarithms to all orders in perturbation theory. 

Factorization of event shape distributions in SCET was prov-en in Refs.~\cite{Fleming:2007qr,Bauer:2008dt}, and factorization for multijet observables defined with arbitrary algorithms was considered in Ref.~\cite{Bauer:2008jx}. The extension to the more general case that we consider involves the straightforward combination of the techniques developed in these papers and will be derived in detail in Ref.~\cite{Ellis:2010rw}. In this work we demonstrate that, after intricate cancellations among the various contributions to the jet and soft functions, consistency of the factorization theorem is satisfied at next-to-leading logarithmic (NLL) accuracy. In order for the factorization theorem to be consistent, the hard, jet, and soft functions defined must satisfy a strong condition on their anomalous dimensions:
\begin{equation}
\label{consistency}
\begin{split}
0 = &\left(\gamma_H + \sum_{i=M+1}^N \gamma_{J_i}\right) \delta(\tau_a^1) \cdots \delta(\tau_a^M) \\
&+ \sum_{i=1}^M\gamma_{J_i}(\tau_a^i) \prod_{\substack{j=1 \\ j\not = i}}^M \delta(\tau_a^j)  + \gamma_S(\tau_a^1,\dots,\tau_a^M)\,,
\end{split}
\end{equation} 
for any number $N$ of total jets and $M$ of measured jets, and any color representation of each jet. This consistency condition is made even more nontrivial by the potential dependence of the jet and soft anomalous dimensions on the jet algorithm parameters.  In this Letter we demonstrate that \eq{consistency} does in fact hold for arbitrary numbers, types, and sizes of jets in the final state, up to certain power corrections we are able to identify.

We begin in \sec{sec:cuts} by defining the phase space cuts needed to implement our choice of jet algorithms.  In \sec{sec:fact} we then present the factorization theorem for $N$-jet events and define the hard, jet, and soft functions, and identify power corrections to the factorization. In \sec{sec:RGE} we give the form of the RG evolution equations obeyed by the factorized functions.  In \sec{sec:anom} we summarize the results of all the anomalous dimensions needed for NLL running and demonstrate how they intricately satisfy the consistency condition \eq{consistency}. This requires calculating only the infinite parts of the bare functions. We give the finite pieces of the jet and soft functions (which are not needed at NLL) in Ref.~\cite{Ellis:2010rw}. In \sec{sec:resum} as an example we calculate quark and gluon angularity jet shapes in 3-jet final states with logarithms of $\tau_a$ resummed to NLL accuracy.

\section{Phase Space Cuts and the Jet Algorithm}
\label{sec:cuts}

Two general categories of jet algorithms, cone algorithms and recombination ($\kt$-type) algorithms, are commonly used to find jets.  For a jet composed of two particles, as in a next-to-leading order description, the phase space constraints implied by each type of algorithm become very simple.  In this work we deal with the common forms of cone and (inclusive) $\kt$-type algorithms; our cone algorithms include the Snowmass and SISCone algorithms, and our recombination algorithms include the inclusive $\kt$, Cambridge-Aachen, and anti-$\kt$ algorithms.  Cone algorithms require each particle to be within an angle $R$ of the jet axis, while recombination algorithms require the angle between the two particles to be within an angle $D$ of each other.  If we label the jet axis as $\vect{n}$ and its constituent particles as 1 and 2, then the algorithm constraints for a two-particle jet are:
\begin{equation}
\label{injetconstraint}
\begin{split}
\textrm{cone type: }&\theta_{1\vect{n}} < R \textrm{ and } \theta_{2\vect{n}} < R \, ,\\
\textrm{$\kt$ type: }&\theta_{12} < D \, .
\end{split}
\end{equation}
For the parts of the jet and soft functions that we give in this work, we find that the functional form is the same for cone-type and $\kt$-type algorithms in terms of the angular parameter $R$ or $D$.  Therefore, we will use the more common $R$ in writing down the jet and soft functions, but we note here that the functional form is the same for $\kt$ with the replacement $R\to D$.

Note that, while all algorithms that we consider fall into one of the two constraints in \eq{injetconstraint} at NLO, at higher orders the various algorithms will behave differently. Without taking this into account, we have no guarantee that we can resum all logarithms of jet algorithm parameters correctly. This is not a problem we solve in this paper.
In this paper, we resum logarithms of  jet observables in the presence of phase space cuts due to an algorithm, demonstrate that the factorization theorem and NLL running are valid and consistent, and identify the power corrections to this statement.

At the hard scale, we match an $N$-leg amplitude in QCD onto an $N$-jet operator in SCET, meaning we must enforce that the number of jets is fixed to be $N$. To enforce that we have no more than $N$ jets, we require that the total energy of particles that do not enter jets to be less than a cutoff $\Lambda$. To enforce that we have at least $N$ jets, we need that pairwise each jet is well separated from every other jet. The requirement of consistency of NLL running will give a quantitative measure  of this separation requiring that $t \gg1$.

\section{Factorized Jet Shapes in $N$-Jet Production}
\label{sec:fact}

The cross section for $e^+ e^-$ annihilation to $N$ jets  at center-of-mass energy $Q$, differential in the jet three-momenta $\vect{P}_i$ of the jets and in the shapes of $M$ of these jets, is given in QCD by
\begin{equation}
\label{QCDcs}
\begin{split}
&\frac{d\sigma}{d\tau_a^1 \cdots d\tau_a^M  d^3\vect{P}_1\cdots d^3\vect{P}_N} \\
&\quad\quad = \frac{1}{2Q^2}\sum_X  (2\pi)^4\delta^4(Q-p_X)\abs{\bra{X}j^\mu(0)\ket{0}L_\mu}^2 \\
&\quad\qquad\times\delta_{n(\mathcal{J}(X))-N} \prod_{i=1}^M \delta(\tau_a^i - \tau_a(J_i)) \prod_{j=1}^N\delta^3(\vect{P}_j - \vect{P}(J_j))\,,
\end{split}
\end{equation}
where $J_{i}$ is the $i$th jet in $X$ identified by the jet algorithm $\mathcal{J}$. The Kronecker delta restricts the sum over states to those that are identified as having $N$ jets by the algorithm. The final state is produced by the QCD current $j^\mu = \bar q \gamma^\mu q$, and $L_\mu$ is the leptonic part of the amplitude for $e^+ e^-\to \gamma^*$. 

To factorize the cross section \eq{QCDcs}, we begin by matching the QCD current $j^\mu$ onto a set of $N$-jet operators in SCET.
These operators are built from quark and gluon jet fields,
\begin{equation}
\label{jetfields}
\chi_n = W_n^\dag\xi_n\,,\quad B_n^\perp = \frac{1}{g}W_n^\dag(\mathcal{P}_\perp + A_n^\perp)W_n\,,
\end{equation}
where $\xi_n,A_n$ are collinear quark and gluon fields in SCET, and $W_n$ is a Wilson line of the $\mathcal{O}(1)$ component $\bn\cdot A_n$ of collinear gluons,
\begin{equation}
W_n(x) = \sum_{\text{perms}}\exp \left[-\frac{g}{\bn\cdot\mathcal{P}}\bn\cdot A_n(x)\right]\,.
\end{equation}
We have made use of the label operator $\mathcal{P}^\mu$ which picks out the large $\mathcal{O}(1)$ $\bn\cdot \tilde p$ and $\mathcal{O}(\lambda)$ $\tilde p_\perp$ components of the label momentum $\tilde p$ of collinear field in SCET. We will not need to construct the $N$-jet operators explicitly, but bases of $2,3,4$  jet operators have been given in \cite{Bauer:2002nz,Bauer:2006mk,Marcantonini:2008qn}, respectively.

To describe an $N$-jet cross section, we construct an effective theory Lagrangian by adding $N$ copies of the collinear Lagrangian in SCET (in $N$ different light-cone directions $n_i$) together with one soft Lagrangian. In each collinear sector, we redefine collinear fields by multiplying by Wilson lines of soft gluons to eliminate the coupling of soft gluons to collinear modes in the leading-order SCET Lagrangian \cite{Bauer:2001yt},
$
\xi_n = Y_n^\dag \xi_n^{(0)}$ and $ A_n = \mathcal{Y}_n A_n^{(0)} \,,
$
where 
\begin{equation}
\label{Yndef}
Y_n(x) = P\exp\left[ig\int_0^\infty ds\,n\cdot A_s(ns+x)\right]\,,
\end{equation}
with $A_s$ in the fundamental representation, and $\mathcal{Y}$ similarly defined but in the adjoint representation.

Performing the above steps in \eq{QCDcs} for the jet shape distribution, the details of which we report in \cite{Ellis:2010rw}, we obtain the factorized form in SCET,
\begin{equation}
\label{SCETcs}
\begin{split}
&\frac{d\sigma}{\prod_{i=1}^M d\tau_a^i \prod_{k=1}^N d^3\vect{P}_k} = \frac{d\sigma^{(0)}}{\prod_{k=1}^Nd^3\vect{P}_k} H(\vect{P}_1,\dots,\vect{P}_N)\!\!\prod_{j=M+1}^{N}\!\! J_{n_j,\omega_j}^{f_j}\\
&\times \prod_{i=1}^M \int\!d\tau_J^i \, d\tau_S^i \, \delta(\tau_a^i - \tau_J^i - \tau_S^i) \,J_{n_i,\omega_i}^{f_i}(\tau_J^i) S(\tau_S^1,\dots,\tau_S^M)\,,
\end{split}
\end{equation}
where $\sigma^{(0)}$ is the Born cross section for $e^+ e^-\to N\text{ partons}$, $H = 1+\mathcal{O}(\alpha_s)$ is the hard coefficient given by the matching coefficient of the SCET $N$-jet operator, and $J$ and $S$ are jet and soft functions. The superscripts $f_{i}$ denote the color representation (corresponding to a quark, antiquark, or gluon) of the jet corresponding to the $i$th leg in the $N$-jet operator. We number the legs so that $i=1,\dots,M$ are the jets whose shapes we measure, and the remainder $j=M+1,\dots,N$ are left unmeasured. 

The quark  and gluon jet functions for jets whose shapes are measured are defined by
\begin{subequations}
\label{jetfuncs}
\begin{align}
\begin{split}
\label{quark}
&J^q_{n,\omega}(\tau_J) =  \frac{1}{\CA} \Tr\sum_{X_n}\int\frac{dn\mcdot k}{2\pi} \int d^4 x \, e^{-ik\cdot x} \frac{\bnslash}{2}  \delta_{n(\mathcal{J}(X_n)) - 1}  \\
&\quad\times\bra{0} \chi_{n,\omega}(x)\ket{X_n}\bra{X_n}\bar\chi_{n,\omega}(0)\ket{0} \delta(\tau_J - \tau_a(J(X_n)))\,,
\end{split} \\
\label{gluon}
\begin{split}
&J^g_{n,\omega}(\tau_J) =  \frac{\omega}{2\CA \CF} \Tr\sum_{X_n}\int\frac{dn\mcdot k}{2\pi} \int d^4 x \, e^{-ik\cdot x}  \delta_{n(\mathcal{J}(X_n)) - 1}  \\
&\times\frac{1}{D-2}\bra{0} gB_{n,\omega}^{\perp\mu}(x)\ket{X_n}\bra{X_n}gB_{n,\omega\mu}^{\perp}(0)\ket{0}\delta(\tau_J - \tau_a(J(X_n)))\,,
\end{split}
\end{align}
\end{subequations}
where the traces are over color and spinor indices, and $D$ is the number of dimensions. The sums are over states in the $n$-collinear sector. The label direction and energy $n,\omega$ are chosen to match the jet momentum $\vect{P}$. We have factored the Kronecker delta in the full cross section \eq{QCDcs} restricting the sum over states to those with $N$ jets according to the algorithm $\mathcal{J}$ into individual restrictions that there is precisely one jet in each collinear sector. The delta functions of $\tau_J$ restrict the angularity of the jet $J$ identified in the state $X_n$ by the jet algorithm. The jet functions $J_{n_j,\omega_j}^{f_j}$ for jets whose shapes are left unmeasured are given by \eq{jetfuncs} without the delta functions of $\tau_J$. 

The soft function, meanwhile, is given by matrix elements of  $N$ soft Wilson lines in each of the collinear directions $n_i$ and color representations $r_i$ of the $i$th jet. For arbitrary $N$, multiple color structures may appear, and if so there is an implicit sum over multiple hard functions $H$ and soft functions $S$ in \eq{SCETcs}. An $N$-jet soft function takes the general form,
\begin{equation}
\label{softfunc}
\begin{split}
S_{N}&(\{\tau_S^i\}) = \frac{1}{\mathcal{N}} \sum_{X_s}\delta_{n(\mathcal{J}(X_s))} \prod_{i=1}^M \delta(\tau_S^i - \tau_a^{i}(X_s)) \\
& \times\bra{0} Y_{n_N}^{r_N\dag}\cdots Y_{n_1}^{r_1\dag}(0)\ket{X_s} \bra{X_s} Y_{n_1}^{r_1} \cdots Y_{n_N}^{r_N}(0)\ket{0}\,,
\end{split}
\end{equation} 
where $\mathcal{N}$ normalizes the soft function to $\delta(\tau_a^1)\cdots\delta(\tau_a^M)$ at tree level. There is an implicit contraction of color indices which we have left unspecified. The whole soft function is color singlet.
Note that the sum over soft states is restricted so that soft particles do not create an additional jet when the jet algorithm is run on $X_s$. $\tau_a^{i}(X_s)$ is the contribution to the jet shape from soft particles which are actually in the jet $J_i$. 

The factorization of the cross section \eq{SCETcs} is valid in the following limits of QCD:
\begin{enumerate}
\item The SCET expansion parameter $\lambda$, determined either by the jet shape $\tau_a$ for measured jets or the jet radius $R$ for unmeasured jets, must be small. In other words, each jet must be \emph{well collimated}.

\item The separation between any pair of jets must be large. We will find that the natural measure for this separation is the variable $t = \tan(\psi/2)/\tan(R/2)$, where $\psi$ is the minimum angle between two jet directions. $t$ must be large, that is, jets must be \emph{well separated} in order for us to factor the $N$-jet condition in the full cross section \eq{QCDcs} into $N$ individual 1-jet conditions in each collinear sector as in \eq{jetfuncs} and a no-jet condition in the soft sector as in \eq{softfunc}. 
This approximation is inevitable because each jet function $J_i$  already approximates all radiation emitted by other jets as coming from a Wilson line $W_{n_i}$ along the exactly back-to-back direction $\bar n_i$, whereas the hard and soft functions  know the directions of all $N$ jets exactly.

\item The energy of all particles not included in a jet must be of the order of soft momenta. This is so that setting the label energy on each of the jet fields in \eq{jetfuncs} to be equal to the total jet energy is correct at leading order in $\lambda$. In particular, the energy cut parameter $\Lambda$ on energy outside of all jets is required to be soft, $\Lambda\sim \lambda^2 E_J$. 

\item Power corrections associated with the jet algorithm are small. For instance, setting the jet axis equal to the label direction $n$ is valid up to $\mathcal{O}(\lambda^2)$ corrections, which induce corrections to the jet shape $\tau_a^J$ which are subleading for $a<1$ \cite{Berger:2003iw,Bauer:2008dt,Lee:2006nr}. Similarly, assuming soft particles know only about the total collinear jet momentum by the time they are included or excluded from a jet  induces power corrections to $\tau_a^J$ that are power suppressed for sufficiently large $R$.
\end{enumerate}
We go into greater detail about these approximations in \cite{Ellis:2010rw}.

\section{Renormalization Group Evolution}
\label{sec:RGE}

The functions that we consider either renormalize multiplicatively or through convolutions in $\tau$. The multiplicative form of a renormalization group equation (RGE) obeyed by a function $F$ is
\begin{equation}
\label{RGEF}
\mu\frac{d}{d\mu}F(\mu) = \gamma_F(\mu)F(\mu)\,,
\end{equation}
with the anomalous dimension of the form
\begin{equation}
\label{gammaF}
\gamma_F(\mu) = \Gamma_F[\alpha] \ln\frac{\mu^2}{\omega^2} + \gamma_F[\alpha]\,.
\end{equation}
This RGE has the solution
\begin{equation}
F(\mu) = U_F(\mu,\mu_0)F(\mu_0)\,,
\end{equation}
where
\begin{equation}
\label{UF}
U_F(\mu,\mu_0) = e^{K_F(\mu,\mu_0)}\left(\frac{\mu_0}{\omega}\right)^{\omega_F(\mu,\mu_0)}\,,
\end{equation}
where we define $\omega_F,K_F$ below in \eq{omegaK}.
The convolved form of an RGE obeyed by functions $F$ that depend on the observable is
\begin{equation}
\label{RGEFtau}
\mu\frac{d}{d\mu}F(\tau;\mu) = \int d\tau' \gamma_{F}(\tau-\tau';\mu) F(\tau';\mu)\,,
\end{equation}
where
\begin{equation}
\label{gammaFtau}
\gamma_{F}(\tau;\mu) = \left(\Gamma_F[\alpha]\ln\!\frac{\mu^2}{\omega^2} + \gamma_F[\alpha]\right)\delta(\tau) - \frac{2}{j_F}\Gamma_F[\alpha]\left[\frac{\theta(\tau)}{\tau}\right]_+ \!\! .
\end{equation}
The solution to this RGE is \cite{Fleming:2007xt,Becher:2006mr,Korchemsky:1993uz,Balzereit:1998yf,Neubert:2005nt}
\begin{equation} 
F(\tau;\mu) = \int d\tau' U_F(\tau-\tau';\mu,\mu_0)F(\tau';\mu_0)\,,
\end{equation}
where
\begin{equation}
\label{UFtau}
U_F(\tau;\mu,\mu_0) = \frac{e^{K_F + \gamma_E\omega_F}}{\Gamma(-\omega_F)} \left(\frac{\mu_0}{\omega}\right)^{j_F\omega_F}\left[\frac{\theta(\tau)}{\tau^{1+\omega_F}}\right]_+\,.
\end{equation}
We note that the anomalous dimensions $\gamma_F(\mu)$ and $\gamma_F(\tau; \mu)$ in general also depend on the jet algorithm parameters $R$ and $\Lambda$ which we have made implicit.

The part of the anomalous dimensions in \eqs{gammaF}{gammaFtau} multiplying $\ln(\mu^2/\omega^2)$ is proportional, to all orders in $\alpha_s$, to the \emph{cusp anomalous dimension} $\Gamma(\alpha_s)$, given to $\mathcal{O}(\alpha_s)$ by $\Gamma(\alpha_s) = \alpha_s/\pi$.  With one-loop results for the anomalous dimensions, and using the two-loop form of the cusp anomalous dimension, the RGE solutions are accurate to NLL order.  In \eqs{UF}{UFtau}, $\omega_F,K_F$ are given by
\begin{subequations}
\label{omegaK}
\begin{align}
\label{omegaF}
\omega_F(\mu,\mu_0) &= \frac{2}{j_F}\int_{\alpha_s(\mu_0)}^{\alpha_s(\mu)}\frac{d\alpha}{\beta[\alpha]}\Gamma_F[\alpha] \\
\label{KF}
\begin{split}
K_F(\mu,\mu_0) &= \int_{\alpha_s(\mu_0)}^{\alpha_s(\mu)}\frac{d\alpha}{\beta[\alpha]}\gamma_F[\alpha] \\
&\quad + 2\int_{\alpha_s(\mu_0)}^{\alpha_s(\mu)}\frac{d\alpha}{\beta[\alpha]}\Gamma_F[\alpha] \int_{\alpha_s(\mu_0)}^{\alpha}\frac{d\alpha}{\beta[\alpha]}\,,
\end{split}
\end{align}
\end{subequations}
where $\beta[\alpha]$ is the beta function of QCD. 
We define $j_F = 1$ for RGEs of the form \eq{gammaF}.

We will find that the hard function can be written as a sum over functions that each obey a multiplicative renormalization group equation.  The unmeasured jet function also obeys a multiplicative RGE, while the measured jet function obeys a RGE with a convolution over $\tau$.  The soft function, whose structure we will discuss in detail, can be decomposed into terms which obey multiplicative RGEs and terms which obey convolved RGEs.  

In the next section we outline the calculations necessary to obtain all the above anomalous dimensions to $\mathcal{O}(\alpha_s)$.

\section{Anomalous Dimensions and Consistency of Factorization}
\label{sec:anom}

In this section we discuss the calculation of the one-loop hard, jet, and soft anomalous dimensions and the form of the anomalous dimensions in Table~\ref{table:gammas} and demonstrate that the consistency condition, \eq{consistency}, is satisfied to one-loop order, to leading order in the approximations we enumerated above. This is already an intricate test whose satisfaction turns out to be highly nontrivial. Having verified this condition, we proceed at the end of the Letter to give an application of NLL resummation of the jet shape distribution made possible by  our one-loop calculation of the anomalous dimensions. 

\subsection{Hard Function}
\label{sec:hard}

The hard function $H$ in the factorized cross section \eq{SCETcs} is given by the square of the Wilson coefficient in the matching of the $N$-parton amplitude in QCD onto an $N$-jet operator in SCET,
\begin{equation}
\bra {N} \bar q \Gamma q\ket{0} = \bra{N}C_N\mathcal{O}_N\ket{0}\,,
\end{equation}
where the right-hand side is actually a sum over many possible $N$-jet operators built from the jet fields in \eq{jetfields} and soft Wilson lines \eq{Yndef}. The allowed basis of operators $\mathcal{O}_N$ is determined by gauge and Lorentz symmetry. If there is only one operator, the hard function is simply $H = \abs{C_N}^2$.

The one-loop anomalous dimension of the $N$-jet matching coefficient $C_N$ can be determined from calculations existing in the literature, for example, Table III of Ref.~\cite{Chiu:2009mg}. For an operator with $N$ legs with color charges $\vect{T}_i$, the anomalous dimension of the matching coefficient $C_N$ is
\begin{equation}
\begin{split}
\gamma_{C_N}  (\alpha_s) = & - \sum_{i=1}^N\left[ \vect{T}_i^2 \Gamma(\alpha_s) \ln\frac{\mu}{\omega_i} + \frac{1}{2}\gamma_i(\alpha_s)\right] \\
& - \frac{1}{2}\Gamma(\alpha_s) \sum_{i\not = j}\vect{T}_i\cdot \vect{T}_j\ln\left(\frac{-n_i\cdot n_j - i0^+}{2}\right)
\end{split}
\end{equation}
where $\gamma_i$ is given to $\mathcal{O}(\alpha_s)$ for quarks and gluons by
\begin{equation}
\label{noncusp}
\gamma_q = \frac{3\alpha_sC_F}{2\pi}\,,\quad \gamma_g = \frac{\alpha_s}{\pi}\frac{11 C_A - 4 T_R n_f}{6}\,.
\end{equation}
The anomalous dimension of the hard function itself is then given by $\gamma_H = \gamma_{C_N} +\gamma_{C_N}^*$ and can be written as
\begin{equation}
\label{hardanomdim}
\gamma_H(\mu) = \sum_{i = 1}^N \gamma_H^i(\mu) + \gamma_H^{\text{pair}}(\mu)\,.
\end{equation}
Because the hard function obeys a multiplicative RGE, each term in the hard function obeys a multiplicative RGE, and so each term in \eq{hardanomdim} has the form \eq{gammaF}. Each $H^i$ has $\omega = \omega_i$, while $\Gamma[\alpha] = 0$ for $H^{\text{pair}}$, as listed in Table~\ref{table:gammas}.

\subsection{Jet Functions}

The quark and gluon jet functions are given by \eqs{quark}{gluon} and are calculated from cutting all possible diagrams at a given order in $\alpha_s$ correcting a collinear propagator  with label momentum $\omega n$.  The jet functions include phase space restrictions on the final-state particles from the cut requiring that only one jet is produced.  When we cut through a single propagator, the solitary parton in the final state is automatically in the jet, but these diagrams turn out to be scaleless and thus zero in dimensional regularization. For the cuts through loops, two collinear particles are created in the final state, and both particles are in the jet if \eq{injetconstraint} is satisfied.  If \eq{injetconstraint} is not satisfied, we require one of the particles to have energy $E < \Lambda$, so that only one jet is produced by the final state.  Additionally, for jets whose shapes are measured, we include a delta function, $\delta(\tau_J - \tau_a(J(X)))$, measuring the jet shape for the particles in the jet.  The restrictions on unmeasured jet functions are the same as the measured jets except for this delta function.

We report here the results of calculating only the infinite parts of the relevant loop graphs in dimensional regularization, in $D = 4-2\epsilon$ dimensions, in the $\overline{\text{MS}}$ scheme. We give the finite parts in \cite{Ellis:2010rw}. Our calculations give anomalous dimensions for quark and gluon jets $\gamma_J^i$ of the form \eq{gammaF} for unmeasured jets and $\gamma_{J}^k(\tau_a)$ of the form \eq{gammaFtau} for measured jets, with the values given in \tab{table:gammas}. 

In the measured jet function, we find that the zero-bin subtraction plays a key role.  The zero-bin subtraction removes doubly-counted regions of phase space from the ``na\"{\i}ve'' contributions to the jet function \cite{Manohar:2006nz}.  For the measured jet functions, the na\"{\i}ve contributions to the anomalous dimension only depend on $\delta(\tau_a)$ and do not contain $(1/\tau_a)_+$ distributions.  However, the zero-bin contribution to the anomalous dimension contains non-trivial $\tau_a$ dependence away from $\tau_a=0$, and it is only by performing the zero-bin subtraction that we obtain the correct running of the measured jet function.  

When the final-state particles in the jet function do not pass the cuts in \eq{injetconstraint}, only one particle is in a jet.  In this case the contribution to the jet function is power suppressed by $\cO(\Lambda/\omega)$, since a collinear parton must have $E<\Lambda$ to be outside of the jet.  This power contribution is not  power suppressed in the na\"{\i}ve contribution alone, but only after the zero-bin subtraction.  Additionally, the zero-bin removes the dependence of the measured jet function anomalous dimension on the jet algorithm parameter $R$.  For unmeasured jets, the zero-bin is a scaleless integral, and the $R$ dependence remains in the unmeasured jet function. 

Tabulating the results, we find the anomalous dimensions are 
\begin{equation}
\label{gammaJ}
\gamma_{J_i} = \Gamma(\alpha_s)\vect{T}_i^2 \ln\frac{\mu^2}{\omega_i^2\tan^2\frac{R}{2}} + \gamma_i\,,
\end{equation}
for unmeasured jet functions, and
\begin{equation}
\label{gammaJM}
\begin{split}
\gamma_{J_i}(\tau_a^i) &= \vect{T}_i^2\left[\Gamma(\alpha_s)\frac{2-a}{1-a}\ln\frac{\mu^2}{\omega_i^2} + \gamma_i\right]\delta(\tau_a^i) \\
&\quad - 2\Gamma(\alpha_s)\vect{T}_i^2 \frac{1}{1-a}\left[\frac{\theta(\tau_a)}{\tau_a}\right]_+
\end{split}
\end{equation}
for measured jet functions.

\begin{table}
\begin{center}
\begin{tabular}{||c | ccccc ||}
\hline\hline
 & \tabruleA $\Gamma_F[\alpha]$ & & $\gamma_F[\alpha]$ & & $j_F$  \\ \hline
\tabruleA $H^i$ & $-\Gamma \vect{T}_i^2$  && $-\gamma_i$ && 1 \\
\tabruleA $H^{\text{pair}}$ & 0  && $ - \Gamma\sum_{i\not=j}\vect{T}_i\cdot \vect{T}_j\ln\frac{ n_i\cdot n_j}{2}$ && 1 \\
\tabruleA $J^i$ & $\Gamma \vect{T}_i^2$ &&$ \gamma_i - \Gamma \vect{T}_i^2\ln\tan^2\frac{R}{2}$ && 1 \\
\tabruleA $J^k(\tau_a^k)$ &$ \Gamma \vect{T}_k^2\frac{2-a}{1-a}$ && $ \gamma_k$ && $2-a$ \\
\tabruleA $S^k(\tau_a^k)$ &$ - \Gamma\vect{T}_k^2 \frac{1}{1-a} $ && 0 && 1 \\
\tabruleA $S^i$ & 0 && $\Gamma\vect{T}_i^2 \ln\tan^2\frac{R}{2}$ && 1 \\
\tabruleA $S^{\text{pair}}$&  0 && $\Gamma\sum_{i\not=j}\vect{T}_i\cdot\vect{T}_j\ln\frac{n_i\cdot n_j}{2}$ && 1 \\
\hline\hline
\end{tabular}
\end{center}
\caption{Anomalous dimensions of hard, jet, and soft functions. The cusp parts $\Gamma_F$ and non-cusp parts $\gamma_F$ of the anomalous dimensions for hard, unmeasured jet, measured jet, and soft functions are given, along with the constant $j_F$ appearing in \eqs{gammaFtau}{omegaF}. $\Gamma$ is the cusp anomalous dimension, given to one-loop by $\Gamma = \alpha_s/\pi$. The pieces $\gamma_{i}$ for quarks and gluons are given by \eq{noncusp}. The three rows for the soft anomalous dimensions are organized to correspond to the three groups of evolution factors given in \eq{softsolution} and are given in the limit $1/t^2\to 0$.}
\label{table:gammas}
\end{table}

\subsection{Soft Function}

\begin{figure}[b]
\begin{center}
\resizebox{\columnwidth}{!}{\includegraphics{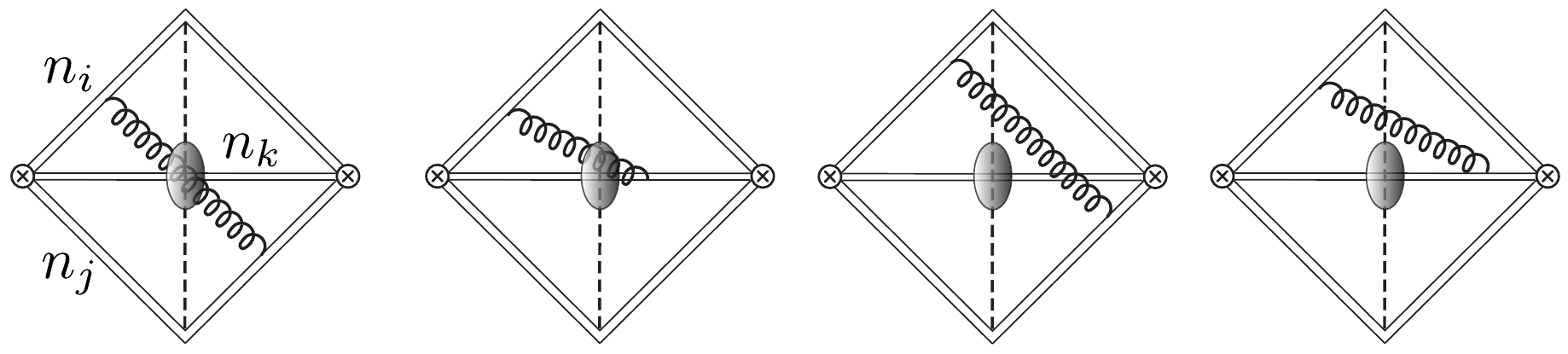}}
\end{center}
\vspace{-2em}
\caption{Soft Function Diagrams. A gluon exchanged between jets $i$ and $j$ crosses the cut which imposes phase space restrictions due to the jet algorithm. The blob represents the jet in direction $k$, which the gluon may enter or not.}
\label{fig:soft}
\end{figure}

The soft function in an $N$-jet cross section is given by  \eq{softfunc}, containing matrix elements of $N$ soft Wilson lines in the $N$ jet directions, with each Wilson line in the color representation of the corresponding jet.  At $\mathcal{O}(\alpha_s)$, this soft function is given by a sum over cut diagrams represented in \fig{fig:soft}.  The blob represents the jet in direction $n_k$, and we leave implicit the phase space cuts needed for each diagram.  We use Feynman gauge, in which each diagram is proportional to $n_i\cdot n_j$. (Note this allows us to drop graphs with $i=j$ or $i=k$ since $n_i^2 = 0$.)  

To calculate the soft function, we must implement phase space cuts on the soft gluon in the final state requiring that it either be in a jet or not produce a new jet (i.e., it has energy less than $\Lambda$).  The soft function is a sum over contributions from all pairs of directions $i$ and $j$ that exchange the soft gluon, and we calculate the total contribution with $i$ and $j$ fixed before summing over directions.  A natural way to organize the phase space of the soft gluon in the final state is as follows:
\begin{enumerate}[(1)]
\item The gluon enters a measured jet and contributes to $\tau_a^{k}(X_s)$. 
\item The gluon enters an unmeasured jet and has any energy. 
\item The gluon is not in any jet and has energy $E<\Lambda$. 
\end{enumerate}
We name contribution (1) $S_{ij}^\text{meas}(\tau_a^k)$, where the subscript $ij$ denotes that the gluon goes from $i$ to $j$. Regions (2) and (3) do not contribute to the angularity of any jet and just give an additive contribution $S_{ij}^{\text{non-meas}}$ to the coefficient of $\delta(\tau_a^1)\cdots \delta(\tau_a^M)$ in the full soft function $S(\tau_a^1,\dots,\tau_a^M)$. Contribution (3), however, is very awkward to calculate, as we must integrate over a phase space with many ``holes'' (corresponding to the jets) removed, resembling Swiss cheese. It is easier to reorganize contributions (2) and (3) into the following form:
\begin{enumerate}
\item[(A)] $S_{ij}^{\text{incl}}$: the gluon is anywhere with energy $E<\Lambda$.
\item[(B)] $S_{ij}^{{k}}$: the gluon is in jet $k$ with energy $E>\Lambda$.
\item[(C)] $\bar S_{ij}^{{k}}$: the gluon is in jet $k$ with energy $E<\Lambda$. 
\end{enumerate}
Then, the unmeasured soft gluon contribution $S_{ij}^{\text{unmeas}}$ (the sum of (2) and (3) in the original list) is given by the combination
\begin{equation}
\label{Snonmeas}
S_{ij}^{\text{unmeas}} = S_{ij}^{\text{\incl}} + \sum_{k=M+1}^N S_{ij}^k - \sum_{k = 1}^M \bar S_{ij}^k \,.
\end{equation}
In the first term, coming from region (A), we filled in the holes in the Swiss cheese-like region (3) in the original list, allowing the soft gluon to go anywhere with energy $E<\Lambda$. We compensated by adding the second term given by region (B) containing gluons  with energy $E>\Lambda$ inside unmeasured jets (part of the original region (2)) and subtracting the third term from region (C), removing gluons with $E<\Lambda$ inside measured jets, which are already correctly accounted for in $S_{ij}^{\text{meas}}(\tau_a^k)$.

The total soft function at $\mathcal{O}(\alpha_s)$ is then given by
\begin{equation}
\label{Stotal}
\begin{split}
S(\tau_a^1,\dots,\tau_a^M) =  \sum_{i\not =j}&\left[ \sum_{k=1}^M S_{ij}^{\text{meas}}(\tau_a^k)\prod_{\substack{l=1 \\ l\not = k}}^M\delta(\tau_a^l) \right. \\
&\quad \left. + S_{ij}^{\text{unmeas}}\prod_{l=1}^M\delta(\tau_a^l) \right] \,.
\end{split}
\end{equation}
Note that the second line is independent of the jet shape. This contribution is universal and will appear in any $N$-jet cross section in which some of the jets defined by a particular jet algorithm are not measured.

The contributions of the measured jet piece $S_{ij}^{\text{meas}}(\tau_a^k)$ to the anomalous dimension of the soft function are given in \tab{table:soft} separately in the cases that $k=i$ or $j$ and $k\not = i,j$.  These contributions are given by the form \eq{gammaFtau}, with the values given in \tab{table:soft}.
The results  are given in terms of the distance measure $t_{ij}=\tan(\psi_{ij}/2)/\tan(R/2)$ between jets of size $R$ separated by an angle $\psi_{ij}$,
and the angle $\beta_{ij}$ between the $ik$ and $jk$ planes. For well-separated jets, the contributions to the non-cusp part of the anomalous dimension are suppressed by $1/t^2$. 

The ``inclusive'' contribution $S_{ij}^{\text{incl}}$ for a soft gluon going anywhere with energy $E<\Lambda$ contributes a term to the soft anomalous dimension given by the general form \eq{gammaF}, with values  given in \tab{table:soft}.

Finally, for the contributions of soft gluons entering jets with $E>\Lambda$  or $E<\Lambda$ in (B) and (C) in the list above, we can combine the last two terms  in \eq{Snonmeas} using the following observation. The sum $S_{ij}^k + \bar S_{ij}^k$ is the contribution of a soft gluon entering jet $k$ with any energy. The phase space integral for this contribution contains a scaleless integral (of energy from 0 to $\infty$), and so this sum is zero in pure dimensional regularization. Thus we can set $\bar S_{ij}^k = -S_{ij}^k$, and the last two terms in \eq{Snonmeas} add up to the contribution of a soft gluon entering \emph{any} jet with energy $E>\Lambda$. 
These contributions can again be split up into those with $k=i$ or $j$ and $k\not=i,j$. They contribute parts to the soft anomalous dimension falling into the form \eq{gammaF}, with values in \tab{table:soft}. The non-cusp pieces are again suppressed by $1/t^2$ for well-separated jets.  

Using the contributions described above, we sum over directions $i$ and $j$ and obtain the anomalous dimensions for $S^{\text{meas}}(\tau_a^k)$ and $S^{\text{unmeas}}$, which we record in \tab{table:soft}.

The soft function obeys the renormalization group equation
\begin{equation}
\label{RGES}
\begin{split}
\mu\frac{d}{d\mu}S(\tau_1,\dots,\tau_M;\mu) = & \int d\tau_1'\cdots d\tau_M'  S(\tau_1',\dots,\tau_M';\mu) \\
&\times \gamma_S(\tau_1-\tau_1',\dots,\tau_M-\tau_M';\mu)\,.
\end{split}
\,,\end{equation}
Because the soft function at $\mathcal{O}(\alpha_s)$ in \eq{Stotal} is a sum of terms that depend non-trivially on at most one jet shape, the anomalous dimension can be decomposed as
\begin{equation}
\label{Sanomdimparts}
\begin{split}
\gamma_S(\tau_1,\dots,\tau_M;\mu) &= \gamma_S^{\text{unmeas}}(\mu)\,\delta(\tau_1)\cdots\delta(\tau_M) \\
& \quad + \sum_{k=1}^M\gamma_S^{\text{meas}}(\tau_k;\mu)\prod_{\substack{j=1 \\ j\not=k}}^M \delta(\tau_j) \,,
\end{split}
\end{equation}
The non-cusp parts of the anomalous dimension of $S^{\text{meas}}$ and $S^{\text{unmeas}}$ share the same dependence on $\tau$, and therefore we are free to shift non-cusp terms freely between anomalous dimensions.  While this does not change the physics, it allows us to organize the anomalous dimensions to match the contributions in \tab{table:gammas}, which we find more convenient for assembling the solution to the soft RGE \eq{RGES}.  By making the non-cusp part of $S^{\text{meas}}(\tau_a^k)$ zero, we find that the shifted $S^{\text{meas}}(\tau_a^k)$ is equal to $S^k(\tau_a^k)$ from \tab{table:gammas}, and that the shifted $S^{\text{unmeas}}$ is equal to $S^{\text{pair}} + \sum_i S^i$.

\begin{table}
\begin{small}
\begin{tabular}{||c | cc @{}c ||}
\hline\hline
 & \tabruleA  $\Gamma_F[\alpha]$ &  $\gamma_F[\alpha]$ &  \\ 
 \hline
\tabruleB   $S_{ij}^{\text{meas}}(\tau_a^i)$ & $\frac{1}{2}\Gamma \vect{T}_i\cdot\vect{T}_j\frac{1}{1-a}$  & $\frac{1}{2}\Gamma\vect{T}_i\cdot\vect{T}_j \ln\frac{t_{ij}^2\tan^2(R/2)}{t_{ij}^2-1} $  &\\
\tabruleB  $S_{ij}^{\text{meas}}(\tau_a^k)$ & 0  & $\frac{1}{2}\Gamma\vect{T}_i\cdot\vect{T}_j \ln\frac{t_{ik}^2t_{jk}^2- 2t_{ik}t_{jk}\cos\beta_{ij}+1}{(t_{ik}^2 - 1)(t_{jk}^2 - 1)}$ &  \\
\tabruleB $S_{ij}^{\text{incl}}$ & $-\Gamma\vect{T}_i\cdot\vect{T}_j$  & $\Gamma\vect{T}_i\cdot\vect{T}_j \,\Big( \ln(n_i\mcdot n_j/2) + \ln \frac{\omega_i^2}{4 \Lambda^2}\Big)$ & \\
\tabruleB $S_{ij}^i$ & $\frac{1}{2}\Gamma\vect{T}_i\cdot\vect{T}_j$  & $-\frac{1}{2}\Gamma\vect{T}_i\cdot\vect{T}_j \, \Big( \ln\frac{t_{ij}^2\tan^2(R/2)}{t_{ij}^2-1} +  \ln \frac{\omega_i^2}{4 \Lambda^2}\Big)$ & \\
\rule{-2pt}{3ex} \rule[-2ex]{0pt}{0pt} $S_{ij}^k$ & 0   & $-\frac{1}{2}\Gamma\vect{T}_i\cdot\vect{T}_j \ln\frac{t_{ik}^2t_{jk}^2- 2t_{ik}t_{jk}\cos\beta_{ij}+1}{(t_{ik}^2 - 1)(t_{jk}^2 - 1)}$   &\\ 
\hline
\rule{-2pt}{3ex} \rule[-1.5ex]{0pt}{0pt} $S^{\text{meas}}(\tau_a^k)$ & $-\Gamma\frac{1}{1-a}\vect{T}_k^2$ & $-\Gamma \vect{T}_k^2\ln\tan^2\frac{R}{2} + \mathcal{O}(1/t^2)$  &\\
\tabruleB $S^{\text{unmeas}}$ & 0 & $\Gamma \sum_{i\not =j}\vect{T}_i\mcdot\vect{T}_j \ln(n_i\cdot n_j/2) $  &\\
\tabruleB & & $+ \Gamma \sum_{i=1}^N \vect{T}_i^2\ln\tan^2(R/2)+ \mathcal{O}(1/t^2)$ & \\
\hline\hline
\end{tabular}
\end{small}
\caption{Soft Anomalous Dimensions. Contributions to the anomalous dimension of the soft function are given for soft gluons emitted by jet $i$ or $j$ and entering jet $k$ (with $k=i$ or $j$ in the first row and $k\not= i,j$ in the second) and being measured with angularity $\tau_a^k$; soft gluons emitted by jet $i$ or $j$ in any direction with energy $E<\Lambda$ in the third row; and soft gluons emitted by jet $i$ or $j$ and entering jet $k$ and angularity unmeasured in the fourth ($k=i$ or $j$) and fifth ($k\not=i,j$) rows. In the second-to-last row we summed the first two rows over all pairs of jets $i,j$ to obtain the measured contribution for a specific $\tau_a^k$, and in the last row, we summed all unmeasured soft gluon contributions. In the last two rows, we have taken the large $t$ limit. $j_F = 1$ in all cases.}
\label{table:soft}
\end{table}

Finally, we can give the soft function anomalous dimension.  Omitting terms which are suppressed by $\mathcal{O}(1/t^2)$, the soft function anomalous dimension is
\begin{equation}
\label{gammaS}
\begin{split}
\gamma_S &(\tau_a^1,\dots,\tau_a^M)  = \Gamma(\alpha_s)\biggl[ - \frac{1}{1-a}\sum_{k=1}^M \vect{T}_k^2\ln\frac{\mu^2}{\omega_k^2} \\
& + \sum_{i=M+1}^N \vect{T}_i^2\ln\tan^2\frac{R}{2} + \sum_{i\not = j}\vect{T}_i\mcdot\vect{T}_j\ln\frac{n_i\cdot n_j}{2}\biggr] \\
&\qquad\times \delta(\tau_a^1)\cdots \delta(\tau_a^M) \\
& + 2\Gamma(\alpha_s)\frac{1}{1-a}\sum_{k=1}^M \vect{T}_k^2 \left[\frac{\theta(\tau_a^k)}{\tau_a^k}\right]_+\prod_{\substack{j=1 \\ j\not=k}}^M\delta(\tau_a^j)\,,
\end{split}
\end{equation}

The solution of the RGE is
\begin{equation}
\label{softsolution}
\begin{split}
&S(\tau_1,\dots,\tau_M;\mu) = \int\!d\tau_1'\cdots d\tau_M' \,S(\tau_1',\dots,\tau_M';\mu_0) \\
&\quad\times   U_S^{\text{pair}}(\mu,\mu_0)\prod_{k=1}^M U_S^k(\tau_k-\tau_k';\mu,\mu_0) \! \prod_{i=M+1}^N\! U_S^i(\mu,\mu_0) \,,
\end{split}
\end{equation}
where $U_S^k(\tau_k)$ is an evolution kernel of a convoluted RGE  and is of the form in \eq{UFtau}, and $U_S^i$ and $U_S^{\text{pair}}$ are evolution kernels of multiplicative RGEs and are of the form in \eq{UF}.  The evolution kernels $U_S^k(\tau_k)$, $U_S^i$, and $U_S^{\text{pair}}$ correspond to the soft anomalous dimensions from $S^k(\tau_a^k)$, $S^i$, and $S^{\text{pair}}$ in \tab{table:gammas}.

\subsection{Consistency of Factorization}
\label{sec:consistency}

Adding together all jet and soft anomalous dimensions, we find, miraculously, the $R$ dependence cancels between the unmeasured jet anomalous dimension \eq{gammaJ} and sum over unmeasured jets in the soft function \eq{gammaS}, and the $\tau_a \not = 0$ dependence cancels between the measured jet anomalous dimension \eq{gammaJM} and the sum over measured jets in the soft function. The remaining pieces precisely match the hard anomalous dimension $\gamma_H$ given in \sec{sec:hard} such that the consistency condition \eq{consistency} is satisfied. Note, however, that satisfying \eq{consistency} exactly required that we drop corrections of $\mathcal{O}(1/t^2)$ in the soft function. Requiring consistency of the anomalous dimensions at one loop has provided the measure $t^2\gg1$  to quantify the condition we used in justifying the factorization theorem in \sec{sec:fact} that jets be ``well separated''.

\section{Application: Jet Shapes in $e^+ e^-\to$ 3 Jets}
\label{sec:resum}

As an example of using the above results to calculate a jet observable in an exclusive multijet final state, we give the resummed angularity jet shape distribution for a single measured quark or gluon jet in a three-jet final state in $e^+ e^-$ annihilation.  The techniques to derive and solve the RGEs to resum logarithms in jet shape distributions in SCET  are essentially identical to those for event shape distributions as performed in 
 \cite{Fleming:2007xt,Hornig:2009vb,Schwartz:2007ib,Becher:2008cf}.
  
We assemble the appropriate RG-evolved hard function, measured  jet function, two unmeasured jet functions, and  soft function given in Secs.~\ref{sec:RGE} and \ref{sec:anom}. Evolving these from their tree-level values at initial scales $\mu_H,\mu_J^i,\mu_S$ to the scale $\mu$ with NLL running, we obtain the distribution in the shape $\tau_a$ of jet 1 with jets 2,3 unmeasured,
\begin{equation}
\label{3jetshape}
\begin{split}
&\frac{1}{\sigma^{(0)}_{\vect{P}_1\vect{P}_2\vect{P_3}}}\frac{d\sigma_{\vect{P}_1\vect{P}_2\vect{P_3}}}{d\tau_a} = \exp\bigl[\mathcal{K}(\mu;\mu_{H},\mu_J^{1,2,3},\mu_S)\bigr] \\
&\qquad \times \frac{\exp\bigl[\gamma_E\bigl( \omega_J^1(\mu,\mu_J^1)  + \omega_S^1(\mu,\mu_S)\bigr)\bigr]}{\Gamma(-\omega_J^1(\mu,\mu_J^1)  - \omega_S^1(\mu,\mu_S)\bigr)} \left(\frac{\mu_H}{\bar\omega_H}\right)^{\omega_H(\mu,\mu_H)} \\
&\qquad\times\left(\frac{\mu_J^1}{\omega_1}\right)^{(2-a)\omega_J^1(\mu,\mu_J^1)}\left(\frac{\mu_J^2}{\omega_2}\right)^{\omega_J^2(\mu,\mu_J^2)} \left(\frac{\mu_J^3}{\omega_3}\right)^{\omega_J^3(\mu,\mu_J^3)} \\
&\qquad\times \left(\frac{\mu_S}{\omega_1}\right)^{\omega_S^1(\mu,\mu_S)}\left[\frac{1}{\tau_a^{1+\omega_J^1(\mu,\mu_J^1)+\omega_S^1(\mu,\mu_S)}}\right]_+\,,
\end{split}
\end{equation}
where $\sigma_{\vect{P}_1\vect{P}_2\vect{P_3}}$ is the cross section differential in the three jet momenta $\vect{P}_i = \omega_i \vect{n}_i$, the effective hard scale $\bar\omega_H = (\omega_1^{\vect{T}_1^2}\omega_2^{\vect{T}_2^2}\omega_3^{\vect{T}_3^2})^{\frac{1}{\vect{T}^2}}$ where $\vect{T}^2 = \vect{T}_1^2 +  \vect{T}_2^2 +  \vect{T}_3^2$, and $\mathcal{K}$ is the sum of the hard, jet, and soft evolution factors,
\begin{equation}
\begin{split}
\mathcal{K}&= K_H(\mu,\mu_H) + \sum_{i=1}^3 [K_J^i(\mu,\mu_J^i) + K_S^i(\mu,\mu_S)] + K_S^{\text{pair}}(\mu,\mu_S)\,.
\end{split}
\end{equation}
Inspection of \eq{3jetshape} suggests the reasonable choices for initial scales 
\begin{equation}
\mu_H = \bar\omega_H , \ \mu_J^1 = \omega_1\tau_a^{1/(2-a)} ,\  \mu_J^{2,3} = \omega_{2,3}\tan\frac{R}{2} , \ \mu_S = \omega_1\tau_a\,.
\end{equation}
For the unmeasured jet scales $\mu_{J}^{2,3}$ we kept in mind the factor of $\ln\tan^2\frac{R}{2}$ present in $K_J^2$ (see \tab{table:gammas}).  To obtain the shape of a quark or gluon jet from \eq{3jetshape} we designate jet 1 as either quark or gluon and plug in the appropriate color factors and anomalous dimensions from \tab{table:gammas} into $\omega_F$ and $K_F$ appearing in \eq{3jetshape}. We report on a more detailed phenomenological study of these jet shapes in \cite{Ellis:2010rw} and their application to the discrimination of quark vs. gluon jets  in future work.

\section{Conclusions}

We have demonstrated the intricate fashion in which the factorized cross section to produce exclusive $N$-jet final states when $M \le N$ are measured with a jet observable remains consistent for NLL running. We identified sources of power corrections to this factorization theorem and the consistency condition. Up to these corrections, the factorization theorem remains consistent independently of the number of measured and unmeasured jets and number of quark and gluon jets.

One novel power correction that explicitly manifested itself in our calculation is in the separation parameter $t$. Since $1/t$ is identically zero for all jet sizes when jets are back-to-back, this parameter has not been identified in the literature before.

We find that, when a jet measurement is performed, the NLL resummed result has no dependence on the jet algorithm across the algorithms we considered (the Snowmass and SISCone cone algorithms and the inclusive $\kt$, anti-$\kt$, and the Cambridge-Aachen $\kt$-type algorithms). In addition, for unmeasured jets the dependence on the jet algorithm parameter $R$ (or $D$) is universal across these algorithms at NLL.

Jet shapes such as angularities can be used to describe the substructure of a jet, and can be used, for instance, to distinguish quark jets from gluon jets. In a future publication we will develop and describe a strategy to do so. We presented our calculations in such a way that  allows for straightforward adaptation to other measurements as well, as we separated those parts of the jet and soft function that depend only on the jet algorithm and not the choice of jet observable. In addition, the ideas we discussed such as the power corrections that arise in the factorization formula and the method of calculating the soft and jet functions, will carry over to a calculation involving jet algorithms at hadron colliders, essentially amounting to having algorithm parameters that are invariant under boosts along the beam axis.

\section*{Acknowledgements}
We are grateful to C. Bauer for valuable discussions and review of the draft. 
The authors at the Berkeley CTP and in the Particle Theory Group at the University of Washington thank one another's groups for hospitality during portions of this work.  
This work was supported in part by the U.S. Department of Energy under Grants DE-FG02-96ER40956 (SDE, CKV, JRW) and DE-AC02-05CH11231 (AH, CL), and by the National Science Foundation under Grant PHY-0457315 (AH, CL). AH was supported in part by an LHC Theory Initiative Graduate Fellowship, NSF grant number PHY-0705682.

\bibliography{shapes}

\end{document}